\documentclass{article}%
\usepackage{eurosym}
\usepackage{amsfonts}
\usepackage{amsmath}
\usepackage{amssymb}
\usepackage{graphicx}%
\providecommand{\U}[1]{\protect\rule{.1in}{.1in}}
\newtheorem{theorem}{Theorem}
\newtheorem{acknowledgement}[theorem]{Acknowledgement}

\begin{document}

\title{Inverse Langevin and Brillouin functions: mathematical properties and physical applications}
\author{Victor Barsan\\IFIN-HH, CIFRA and HHF, Bucharest - Magurele, Romania}
\maketitle

\begin{abstract}
This paper gives a coherent and comprehensive review of the results concerning
the inverse Langevin $L\left(  x\right)  $ and Brillouin functions
$B_{J}\left(  x\right)  $ and the inverse of $L\left(  x\right)  /x$ and
$B_{J}\left(  x\right)  /x$. As these functions are used in several fields of
physics, without evident interconnections - magnetism (ferromagnetism,
superparamagnetism, nanomagnetism, hysteretic physics), rubber elasticity,
rheology, solar energy conversion - the new results are not always efficiently
transferred from a domain to another. The increasing accuracy of experimental
investigations claims an increasing accuracy in the knowledge of these
functions, so it is important to compare the accuracy of various approximants
and even to obtain, in some cases, the exact form of the inverses of $L\left(
x\right)  $, $B_{J}\left(  x\right)  ,$ $L\left(  x\right)  /x$ and$\ B_{J}%
\left(  x\right)  /x.$ This exact form can be obtained, in some cases, at
least in principle, using the recently developed theory of generalized Lambert
functions; in some particular - and also relevant - cases, explicit
expressions for these new special functions are obtained. The paper contains
also some new results, concerning both exact and approximate forms of the
aforementioned inverse functions.

\end{abstract}

\section{Introduction}

In recent years, as a result of increasing accuracy of experimental research
in several domains of physics - superparamagnetism, rubber elasticity,
hysteretic physics, ferromagnetism - and of significant progress registered in
both numerical and theoretical study of several transcendental equations, much
effort was invested in the investigation of inverse Langevin and Brillouin
functions. A number of approximatns, of several precisions or degrees of
sophistication, exists, and a number of exact solutions are proposed. A
comprehensive comparative study of the approximants, of their accuracy and
applicability, and of their connections with the exact results - when they are
available - is missing, and the main goal of this paper it to fill this gap.
Some original results are also presented. Our intention was to give a coherent
presentation of the status of the art in this domain, of interest for
experimentalists, theorists and mathematicians.

The structure of this paper is the following. In Section 2, the physical
relevance of Brillouin and Langevin functions is discussed, and the importance
of their inverses is illustrated, by two simple examples. The applications of
direct and inverse Langevin and Brillouin functions, which are rarely analyzed
together, are given in the next section. As we shall see, if the Brillouin
function is directly linked to the magnetization of an ideal paramagnet, the
inverse of $B_{J}/x $ is linked to the magnetization of a Weiss ferromagnet.
The main properties of direct and inverse Langevin and Brillouin functions -
their asymptotic behavior, or for small values of arguments, the algebraic
form of Brillouin functions, etc. - are presented in Section 4. Exact results
concerning $L^{-1},\ B_{J}^{-1}$ and related functions, like the inverse of
$B_{J}/x,$ are exposed in Section 5. In the next one, the approximants
of$\ B_{J}^{-1}$ proposed by several authors are critically discussed, and a
recent formula, more accurate than those given till now in literature, is
presented. The last section is devoted to conclusions.

\section{Physical relevance of Brillouin and Langevin functions}

The Langevin and Brillouin functions where introduced in science in the
context of paramagnetism and ferromagnetism. This is why we shall start our
review with an analysis of the simple case of a perfect paramagnet - a system
of $N$ non-interacting quantum magnetic moments $\overrightarrow{\mu}$ in an
external magnetic field $\overrightarrow{B}$ at temperature $T$
\cite{[Suzuki]}. If the magnetic field $\overrightarrow{B}$\ is paralel to the
$Oz $ axis, the magnetic moment has only one non-zero component:
\begin{equation}
\mu_{z}=\overline{\mu}J_{z}\label{1}%
\end{equation}
where the previous equation relates the eigenvalues of the magnetic moment and
of the angular momentum (the relation between the corresponding operators is,
evidently, perfectly similar), and $\overline{\mu}$ is the product of the
Land\'{e} factor $g_{L}$ and Bohr magneton $\mu_{B}:$%

\begin{equation}
\overline{\mu}=g_{J}\mu_{B}.\label{2}%
\end{equation}

The eigenvalues of the energy of a magnetic moment \ are:%

\begin{equation}
\varepsilon\left(  J_{z}\right)  =\overrightarrow{\mu}\overrightarrow{B}%
=\mu_{z}B=\overline{\mu}J_{z}B\label{3}%
\end{equation}

So, the magnetization of the perfect paramagnet is:%

\begin{equation}
M=N\left\langle \mu_{z}\right\rangle =N\overline{\mu}\left\langle
J_{z}\right\rangle =N\overline{\mu}\frac{\sum_{J_{z}=-J}^{J}J_{z}\exp\left(
-\beta\varepsilon\left(  J_{z}\right)  \right)  }{\sum_{J_{z}=-J}^{J}%
\exp\left(  -\beta\varepsilon\left(  J_{z}\right)  \right)  }B\label{4}%
\end{equation}

Let us write the exponent in a more compact form:%

\begin{equation}
\beta\varepsilon\left(  J_{z}\right)  =\beta\overline{\mu}J_{z}B=\frac
{\overline{\mu}BJ}{k_{B}T}\frac{J_{z}}{J}=x\frac{J_{z}}{J}\label{5}%
\end{equation}
where we used the notation:%

\begin{equation}
x=\frac{\overline{\mu}BJ}{k_{B}T}\label{6}%
\end{equation}
With (\ref{5}) and (\ref{6}), eq. (\ref{4}) becomes:%

\begin{equation}
M=N\overline{\mu}BJ\cdot B_{J}\left(  x\right)  =M_{0}\cdot B_{J}\left(
x\right) \label{7}%
\end{equation}
where $B_{J}\left(  x\right)  $ is the Brilloiun function, defined as:%

\begin{equation}
B_{J}\left(  x\right)  =\frac{2J+1}{2J}\coth\left(  \frac{2J+1}{2J}x\right)
-\frac{1}{2J}\coth\left(  \frac{1}{2J}x\right) \label{8}%
\end{equation}
and $M_{0}$ is the maximum value of the magnetization, reached, for instance,
at large fields $\left(  B\rightarrow\infty\right)  $ or small temperatures
$\left(  T\rightarrow0\right)  ,$ when all the magnetic moments become
parralel to the external field. So, we can conclude from (\ref{7}) that the
Brillouin function is the thermal average of $\cos\theta,\ $with $\theta$ -
the angle between $\overrightarrow{\mu}$ and $\overrightarrow{B}. $

In the classical limit, $J\rightarrow\infty$ and the corresponding function,
$B_{\infty}\left(  x\right)  ,$ is called Langevin function:%

\begin{equation}
L\left(  x\right)  =\lim_{J\rightarrow\infty}B_{J}\left(  x\right)  =\coth
x-\frac{1}{x}\label{9}%
\end{equation}

For a classical magnetic moment in an external field, the Zeeman energy is:%

\begin{equation}
U=-\overrightarrow{\mu}\overrightarrow{B}=-\mu B\cos\theta\ \label{10}%
\end{equation}
and the magnetization at thermal equilibrium is proportional to the thermal
average of $\cos\theta:$%

\begin{equation}
\left\langle \cos\theta\right\rangle =\frac{\int e^{-\beta U}\cos\theta
d\Omega}{\int e^{-\beta U}d\Omega}=L\left(  X\right)  \ \label{11}%
\end{equation}
with $X$ defined as:%

\begin{equation}
X=\frac{\mu B}{k_{B}T}.\label{12}%
\end{equation}

With some caution requested by the fact that the average in (\ref{4}),
respectively (\ref{11})\ is taken in quantum, respectively classical context,
it is clear that both $L\left(  x\right)  $ and $B_{J}\left(  x\right)  $
characterize the alignment of a magnetic moment in an external field. So, both
$L\left(  x\right)  $ and $B_{J}\left(  x\right)  ,$ being thermal averages of
$\left\langle \cos\theta\right\rangle ,$ $0<\theta<\pi/2,$ have quite similar
shapes, as functions of $x:$ they are monotonically increasing functions,
starting from zero, and reaching asymptotically the value $1.$ Also, each of
them is equal to the relative magnetization:%

\begin{equation}
m=\frac{M}{M_{0}}\label{13}%
\end{equation}
of the perfect paramagnet. For instance, according to (\ref{7}),%

\begin{equation}
m=B_{J}\left(  x\right) \label{14}%
\end{equation}

Let us consider now a system of interacting quantum magnetic moments, in the
mean-field approximation. In this case, the magnetization is given by a
formula somewhat similar to eq. (\ref{7}), but more complex:%

\begin{equation}
\frac{M\left(  T,H\right)  }{M\left(  0,0\right)  }=\frac{M}{M_{0}}%
=B_{J}\left(  \frac{g_{J}\mu_{B}}{k_{B}T}\left(  H+\lambda M\right)  \right)
\label{15}%
\end{equation}
(see \cite{[Stanley]}, eq. (6.14)), where $\lambda$ is the Curie constant. If
we define the critical temperature $T_{c}$, the maximum value of the
spontaneous magnetization $M_{0}:$%

\begin{equation}
T_{c}=\lambda\frac{N\overline{\mu}^{2}}{4k_{B}},\ M_{0}=N\overline{\mu
}S\ \label{16}%
\end{equation}
and the reduced quantities:%

\begin{equation}
t=\frac{T}{T_{c}},\ h=\frac{\overline{\mu}H}{2k_{B}T_{c}},\ m=\frac{M}{M_{0}%
}\label{17}%
\end{equation}
then eq. (\ref{15}) can be written as:%

\begin{equation}
m\left(  t,h\right)  =B_{J}\left(  \frac{m\left(  t,h\right)  +h}{t}\right)
\label{18}%
\end{equation}
This is the simplest form of the equation of state of the ferromagnet.
Hopefully, the fact that we used the same notation for the magnetization (and
for the reduced magnetization) of a perfect paramagnet, (\ref{7}) and
(\ref{13}), and of a ferromagnet, (\ref{15}), (\ref{17}), will not generate
confusions; of course, these quantities have completely different mathematical expressions.

Putting, in (\ref{18}),%

\begin{equation}
m\left(  t,0\right)  =m\left(  t\right)  \label{19}%
\end{equation}
the equation of state becomes, for $h=0$:%

\begin{equation}
m\left(  t\right)  =B_{J}\left(  \frac{m\left(  t\right)  }{t}\right)
\label{20}%
\end{equation}

Now, we can illustrate the usefulness of the inverses of the functions
$B_{J}\left(  x\right)  $ and $B_{J}\left(  x\right)  /x.$ Taking $B_{J}^{-1}$
in both sides of eq. (\ref{20}), we get:%

\begin{equation}
\frac{B_{J}^{-1}\left(  m\right)  }{m}=\frac{1}{t}\ \label{21}%
\end{equation}
so, we can separate the thermodynamic variables $m$ and $t$ in the equation of
state (see also \cite{[Arrott]}, eq. (12)). However, we cannot obtain, in this
way, an explicit equation for the function $m\left(  t\right)  .$

If we put:%

\begin{equation}
\frac{m\left(  t\right)  }{t}=\zeta\left(  t\right) \label{22}%
\end{equation}
eq. (\ref{20}) can be written as:%

\begin{equation}
t=\frac{B_{J}\left(  \zeta\right)  }{\zeta}\label{23}%
\end{equation}
Denoting by $\beta_{J}\left(  x\right)  $ the inverse of the function
$B_{J}\left(  x\right)  /x,$ we get from (\ref{22}), (\ref{23}):%

\begin{equation}
m\left(  t\right)  =t\beta_{J}\left(  t\right) \label{24}%
\end{equation}
which gives explicitly the dependence of magnetization on temperature - a
relation very useful for experimentalists.

In the limit $J\rightarrow\infty,$ the equation of state (\ref{20}) becomes:%

\begin{equation}
m\left(  t\right)  =L\left(  \frac{m\left(  t\right)  }{t}\right) \label{25}%
\end{equation}
at the variables can be separated, similarly to eq. (\ref{21}), using the
inverse Langevin function $L^{-1}$:%

\begin{equation}
\frac{L^{-1}\left(  m\right)  }{m}=\frac{1}{t}\label{26}%
\end{equation}
However, to get the explicit expression of the magnetization, $m\left(
t\right)  ,$ we need the explicit form of $\lambda\left(  x\right)  ,$ the
inverse of the function $L\left(  x\right)  /x.$ With $\lambda\left(
x\right)  $, this expression is given by a formula similar to (\ref{24}):%

\begin{equation}
m\left(  t\right)  =t\lambda\left(  t\right) \label{27}%
\end{equation}

This is a simple example illustrating the usefulness of the inverses of
functions $B_{J}\left(  x\right)  ,\ L\left(  x\right)  ,\ B_{J}\left(
x\right)  /x,\ L\left(  x\right)  /x$ in magnetism. However, their
applications are not limited to this domain, as we shall see in the next section.

We shall finish this section with a terminological remark. It is interesting
to mention that the "Brillouin functions" were actually introduced in physics
by Debye and by other authors, in the context of old quantum theory, see
\cite{[Stoner]}, \cite{[Van]}. Brillouin used this function in the context of
quantum mechanics. As the term "Debye function" already existed, in the theory
of specific heat of solids, this function took Brillouin's name. However, some
prominent physicists, like Wannier, call both both $L\left(  x\right)  $ and
$B_{J}\left(  x\right)  $ "Langevin functions".

\section{Applications of direct and inverse Langevin and Brillouin functions}

In the mean field theory of ferromagnetism, the Brillouin functions $B_{J}%
$\ are essential ingredients, and the knowledge of the inverse functions
$B_{J}^{-1}$\ allows the exact and explicit calculation of all the
thermodynamic functions of a Weiss ferromagnet \cite{[Polonezii]}, as the
previous example, illustrated by eq. (\ref{21}), suggests. This is not only a
result of theoretical physics, but a starting point of precise determination
of the critical temperature of a ferromagnet, as discussed in \cite{[VB-VK]}.

Similarly to the Weiss model, $B_{J}$ enters also in the equation of state of
arbitrary infinite-range spin Hamiltonians \cite{[KK-SSC-1984]}; so, all the
considerations made for $B_{J}$\ in the context of a Weiss ferromagnet, remain
valid for these systems. The inverse Brillouin functions are also important in
non-iterative mean-field theories, in the determination of the magnetic
entropy \cite{[Katriel-Cont-Fraction-1987]} and in the theory of helical spin
ordering (see \cite{[SSP1967]}, p.332).

They are prominent in the theory of hysteretic phenomena \cite{[Takacs2001]},
\cite{[Takacs2016]}, including the mean-field positive-feedback (PFB) theory
of ferromagnetism [Harrison-JApplPhys]; in the renormalization group theory of
quantum spin systems [Krieg-renorm-gr]. Several recent reviews are devoted to
the inverse Brillouin functions \cite{[Kroger]}, \cite{[Erevan]},
\cite{[VB-review]}.

The Langevin function is important in superparamagnetism
\cite{[super-para1959]} and nanomagnetism - for instance in the theory of
tunnelling magnetoresistance in granular manganites, as the core of
nanoparticles is superparamagnetic, see \cite{[9Jirak]}, especially eqs.
(1-5), and ferrofluids, where the monodomain nanoparticles may also be
super-paramagnetic, see \cite{[Kuncser]}.

The Langevin and Brillouin functions, as well as their inverses, are also
important to polymer science (polymer deformation and flow) and rubber
elasticity. An idealized model for the rubber-like chain is given in Kubo's
treatise of statistical mechanics \cite{[Kubo]}, and is illustrative for
understanding how the Langevin function is used in this domain.

With the development of very precise experimental techniques in the last 30
years, like single-molecule force spectroscopy (SMFS), the elastic strain
energy and the force-displacement relationship, which can be written in terms
of $L^{-1}$ \cite{[1Jedynak]}, \cite{[2Petrosyan]}, could be measured very
precisely; to keep pace with SMFS, the theory had to produce more and more
precise expressions for the inverse Langevin function \cite{[JedynakUkr2018]},
\cite{[JedynakMMS2018]}, \cite{[Morovati]}.

The Langevin function and its inverse are central to truncated exponential
distributions, with applications in solar energy conversion and in many other
domains, including distributions of earthquakes, of forest-fire sizes,
raindrop sizes, reliability modelling, etc., see ref. [1, 3, 6] of
\cite{[12Keady]}. Actually, Keady noticed that the daily clearness index, a
quantity important in solar energy conversion, can be expressed in terms of
the Langevin function \cite{[12Keady]}, \cite{[VB-poster]}.

\section{Basic properties of direct and inverse Langevin and Brillouin
functions}

As previously mentioned,%

\begin{equation}
-1<L\left(  x\right)  <1,\ \ -1<B_{J}\left(  x\right)  <1\label{28}%
\end{equation}
and both functions tend asymptotically to $1.\ $As they are odd:%

\begin{equation}
L\left(  -x\right)  =-L\left(  x\right)  ,\ \ B_{J}\left(  -x\right)
=-B_{J}\left(  x\right) \label{29}%
\end{equation}
and%

\begin{equation}
0\leqslant L\left(  x\right)  ,\ B_{J}\left(  x\right)  <1,\ \ x\geqslant
0\label{30}%
\end{equation}
all the information concerning these functions is contained in the first quadrant.

Near the origin, the Langevin function is:%

\begin{equation}
L\left(  x\right)  =\frac{x}{3}+\mathcal{O}\left(  x^{3}\right) \label{31}%
\end{equation}

For the inverse Langevin function: near the origin,%

\begin{equation}
L^{-1}\left(  x\right)  =3x+\mathcal{O}\left(  x^{3}\right) \label{32}%
\end{equation}
and asymptotically:%

\begin{equation}
L^{-1}\left(  x\rightarrow1\right)  \rightarrow\frac{1}{1-x}\label{33}%
\end{equation}

The Brillouin functions (\ref{8}) can be written also as:%

\begin{equation}
B_{J}\left(  x\right)  =\left(  1+\varepsilon\right)  \coth\left(
1+\varepsilon\right)  x-\varepsilon\coth\varepsilon x,\ \varepsilon=\frac
{1}{2J}\label{34}%
\end{equation}
A useful relation is:%

\begin{equation}
B_{J}\left(  x\right)  =\frac{d}{dx}\ln\sum_{s=-J}^{J}\exp\left(  -\frac
{sx}{J}\right)  =\frac{d}{dx}\ln S_{J}\left(  x\right) \label{35}%
\end{equation}
where the function $S_{J}\left(  x\right)  $ is:%

\begin{equation}
S_{J}\left(  x\right)  =\sum_{s=-J}^{-J+1,...J-1,J}\exp\left(  -\frac{sx}%
{J}\right) \label{36}%
\end{equation}
It is interesting to notice that, as%

\begin{equation}
\int_{0}^{x}B_{J}\left(  y\right)  dy=\ln S_{J}\left(  x\right)  ,\label{37}%
\end{equation}
the integral of the Brillouin function is simpler than the function itself.
Consequently, $S_{J}^{-1}\left(  x\right)  $ can be obtained easier than
$B_{J}^{-1}$, and, if we really obtain it, and if we can write (\ref{37}) in
terms of inverse functions, this could provide us an alternative way of
calculating $B_{J}^{-1}$. This approach has not been explored yet, although
several interesting relations between the integrals of direct and inverse
functions were given in literature \cite{[Kroger]}, \cite{[1Jedynak]}.

Near the origin:%

\begin{equation}
B_{J}\left(  x\right)  =\frac{J+1}{3J}x++\mathcal{O}\left(  x^{3}\right)
\label{38}%
\end{equation}
and asymptotically \cite{[Kroger]}:%

\begin{equation}
B_{J}\left(  x\right)  =1-\frac{1}{J\left(  1+e^{x/J}\right)  }%
,\ \ x\rightarrow\infty\label{39}%
\end{equation}

With a change of variable:%

\begin{equation}
z=e^{\frac{x}{J}}\Leftrightarrow x=J\ln z\ \label{40}%
\end{equation}
both $B_{J}\left(  x\right)  $\ and $S_{J}\left(  x\right)  $\ become
algebraic functions of $z$:%

\[
B_{J}\left(  x\right)  =B_{J}\left(  J\ln z\right)  =
\]

\begin{equation}
=\frac{1}{2J}\frac{\left(  1+2J\right)  \left(  z^{1+2J}+1\right)  \left(
z-1\right)  -\left(  z^{1+2J}-1\right)  \left(  z+1\right)  }{\left(
z^{1+2J}-1\right)  \left(  z-1\right)  }=\overline{B}_{J}\left(  z\right)
=t\ \label{41}%
\end{equation}

\begin{equation}
S_{J}\left(  x\right)  =S_{J}\left(  J\ln z\right)  =z^{J}+z^{J-1}%
+...+z^{-\left(  J-1\right)  }+z^{-J}=\overline{S}_{J}\left(  z\right)
=t\label{42}%
\end{equation}

The roots $z_{J}\left(  t\right)  $\ of the algebraic equation%

\begin{equation}
\overline{B}_{J}\left(  z_{J}\left(  t\right)  \right)  =t\label{43}%
\end{equation}
satisfy also the identity:%

\begin{equation}
B_{J}\left(  J\ln z_{J}\left(  t\right)  \right)  =t\label{44}%
\end{equation}
Applying in both sides of (\ref{44}) the inverse of $B_{J}:$%

\begin{equation}
J\ln z_{J}\left(  t\right)  =B_{J}^{-1}\left(  t\right) \label{45}%
\end{equation}
So, according to (\ref{40}), the quantity $x_{J}\left(  t\right)  ,$ defined as%

\begin{equation}
x_{J}\left(  t\right)  =J\ln z_{J}\left(  t\right) \label{46}%
\end{equation}
where $z_{J}\left(  t\right)  $\ is a convenient root of (\ref{41}), is the
inverse of $B_{J}:$%

\begin{equation}
B_{J}^{-1}\left(  t\right)  =x_{J}\left(  t\right)  =J\ln z_{J}\left(
t\right)  \label{47}%
\end{equation}
The meaning of the term "convenient" will be explined in Subsection 5.2, see
the comments subsequent to eq. (\ref{95}).

Similarily, the convenient root $\widetilde{z}_{J}\left(  t\right)  $ of the
algebraic equation%

\begin{equation}
\overline{S}_{J}\left(  \widetilde{z}_{J}\left(  t\right)  \right)
=t\label{48}%
\end{equation}
gives the inverse of $S_{J}:$%

\begin{equation}
\widetilde{x}_{J}\left(  t\right)  =J\ln\widetilde{z}_{J}\left(  t\right)
=S_{J}^{-1}\left(  t\right) \label{49}%
\end{equation}

So, obtaining $z_{J}\left(  t\right)  $ is equivalent to obtaining $B_{J}%
^{-1}\left(  t\right)  ,$ and obtaining $\widetilde{z}_{J}\left(  t\right)  $
is equivalent to obtaining $S_{J}^{-1}\left(  t\right)  .$

Even if the algebraic form of Brillouin functions is the key for obtaining
their inverses, it is worth mentioning Katriel's approach of obtaining
$B_{J}^{-1}\left(  t\right)  $ as a continued-fraction
\cite{[Katriel-Cont-Fraction-1987]}.

Near the origin, $B_{J}^{-1}\left(  t\right)  $ behaves like:%

\begin{equation}
B_{J}^{-1}\left(  x\right)  =\frac{3J}{J+1}x++\mathcal{O}\left(  x^{3}\right)
\label{50}%
\end{equation}
and asymptotically:%

\begin{equation}
B_{J}^{-1}\left(  t\rightarrow1\right)  \rightarrow J\ln\frac{1}%
{1-t}\label{51}%
\end{equation}
as as we shall see later on, in Subsection 5.2. Kr\"{o}ger [17] obtains the
same singularity, using a different approach, but imposing the symmetry
condition on $B_{J}^{-1}$ (it is an odd function of $t$), which replaces
(\ref{51}) by:%

\begin{equation}
B_{J}^{-1}\left(  t\rightarrow1\right)  \rightarrow2J\tanh^{-1}\left(
t\right)  ,\ \ \ y\rightarrow1\label{52}%
\end{equation}
where the inverse hyperbolic function is, evidently:
\begin{equation}
\tanh^{-1}\left(  t\right)  =\frac{1}{2}\ln\frac{1+t}{1-t}\label{53}%
\end{equation}

In order to obtain explicit expressions for $z\left(  t\right)  ,$
$\widetilde{z}\left(  t\right)  ,$ it is convenient to discuss the case of
integer and half-integer spin separately. We shall do this in Subsection 5.2.

\begin{figure}[ptb]
\begin{center}
\includegraphics[width=\textwidth]{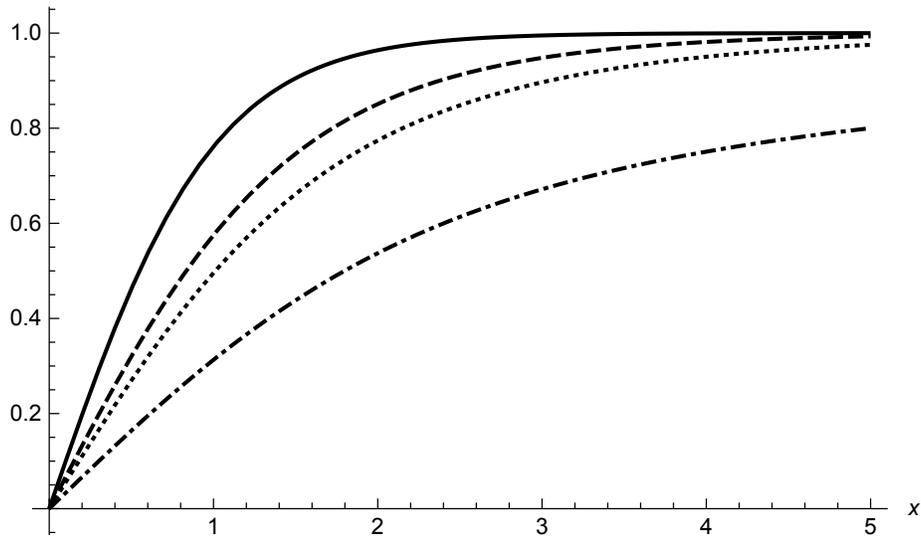}
\end{center}
\caption{The plot of Brillouin functions of indices 1/2 (solid), 1 (dash), 3/2
(dots) and of Langevin function (dot-dash) }%
\end{figure}

As mentioned in Section 2, and visualized in Fig. 1, the Brillouin and
Langevin functions have similar shapes, and this is due to the fact that they
all are a measure of the tendency of an external field to align a magnetic
momentum along its direction. It is clear from Fig. 1 (and easy to prove
mathematically) that%

\begin{equation}
B_{1/2}\left(  x\right)  >B_{1}\left(  x\right)  >...>L\left(  x\right)
\label{54}%
\end{equation}

These inequalities reflect the fact that a smaller spin can be easier aligned
by a certain external field, at a certain temperature, the limiting cases
being the smallest $\left(  J=1/2\right)  $ and the largest $\left(
J=\infty\right)  $\ spin. However, the Brillouin and Langevin functions are
qualitatively different: for instance, they can be inverted by solving an
algebraic equation (Brillouin) or a transcendental one (Langevin), and their
integral on their domain of definition is convergent, respectively divergent.
If we want to speculate, we could notice that the quantum approach is simpler
than the classical one.

\section{Exact results concerning the inverses of $L\left(  x\right)
,\ L\left(  x\right)  /x,\ B_{J}\left(  x\right)  $ and $B_{J}\left(
x\right)  /x$}

\subsection{Generalized Lambert functions approach}

Let us remind that the Lambert function $W\left(  a\right)  $ is the solution
of the transcendental equation:%

\begin{equation}
xe^{x}=a\ \rightarrow x=W\left(  a\right) \label{55}%
\end{equation}
To cite Corless \cite{[CorlessW]} , $W$ "is the simplest example of a root of
an exponential polynomial", i.e. of an expression of the form:%

\begin{equation}
\Pi_{m}\left(  x\right)  =\sum_{i=1}^{m}P_{i}\left(  x\right)  e^{w_{i}%
x}\label{56}%
\end{equation}
with $P_{i}\left(  x\right)  $ - polynomials in $x$, "and exponential
polynomials are the next simplest class of functions after polynomials."
However, there is not much comfort in this simplicity. When the r.h.s. of the
eq. (\ref{56}) contains only one exponential, i.e. $w_{1}=0,\ w_{2}%
\neq0,\ P_{1}\neq0,\ P_{2}\neq0$ and all $P_{i}\equiv0$ for $i>2,$ the
solutions of eq. $\Pi_{m}\left(  x\right)  =0,$ for low indices, are the
generalized Lambert functions, introduced and studied by Mez\"{o}, Baricz
\cite{[Mezo]} and Mugnaini \cite{[Mugnaini]}. Actually, the transcendental
equation $\Pi_{2}\left(  x\right)  =0$ was written in the form \cite{[Mezo]}:%

\begin{equation}
e^{x}\frac{\left(  x-t_{1}\right)  ...\left(  x-t_{n}\right)  }{\left(
x-s_{1}\right)  ...\left(  x-s_{m}\right)  }=a\label{57}%
\end{equation}
where the parameters $t_{1},...,t_{n},s_{1}...,s_{m}$ are supposed to be real.
Its solution or, in other words, the generalized Lambert function, is
denoted:
\begin{equation}
x=W\left(  t_{1},t_{2},...t_{n};s_{1},s_{2},...s_{m};a\right)  .\label{58}%
\end{equation}

The parameters $t_{1},\ t_{2},...$ are called upper parameters, and
$s_{1},s_{2},...$ - lower parameters. Mez\"{o}, Baricz \cite{[Mezo]} and
Mugnaini \cite{[Mugnaini]} obtained compact formulas - relative simple series
expansions - \textit{inter alia}, for the functions $W\left(  \tau
;\sigma;a\right)  .$ However, if the exponential polynomial in (\ref{56})
contains two or more different exponentials, there is no formula for its roots
(to the best of author's knowledge).

There is an interesting connection between $W\left(  \tau;\sigma;a\right)  $
and the Laguerre polynomials:%

\begin{equation}
W\left(  \tau;\sigma;a\right)  =\tau-\left(  \tau-\sigma\right)  \sum
_{n=1}^{\infty}\frac{L_{n}^{\prime}\left(  n\left(  \tau-\sigma\right)
\right)  }{n}e^{-n\tau}a^{n}\label{59}%
\end{equation}
where $\sigma\neq\tau,$ and $L_{n}^{\prime}$ is the first derivative of the
$n-$th order Laguerre polynomial.

As Mez\"{o} and Baricz noticed, the term of "generalized Lambert functions",
for the functions defined by eqs. (\ref{57}), (\ref{58}), is not very
appropriate, as the Lambert $W$ function cannot be obtained, for any
particular choice of the parameters entering in (\ref{57}). However, the
inverse of the function%

\begin{equation}
xe^{x}+rx\label{60}%
\end{equation}
with $r$ - a fixed real number, denoted by $W_{r}\ ,\ $has this property: if
$r=0,$ $W_{r}$ becomes the $W$ Lambert function. The connection between
$W\left(  \tau;\sigma;a\right) $ and $W_{r}$ is (Theorem 3 of \cite{[Mezo]}):%

\begin{equation}
W\left(  \tau;\sigma;a\right)  =\tau+W_{-a\exp\left(  -\tau\right)  }\left(
a\left(  \tau-\sigma\right)  e^{-\tau}\right) \label{61}%
\end{equation}

Among the branches of $W_{r}$, a special role is played by $W_{1/e^{2}}\left(
x\right)  .\ $For $x=-4/e^{2},\ W_{1/e^{2}}\left(  x\right)  $ has a unique
property: it is continuos (as everywhere on the real line) but is not
differentiable, and%

\begin{equation}
W_{1/e^{2}}\left(  -4/e^{2}\right)  =-2\label{62}%
\end{equation}
As we shall see later, it corresponds to the spontaneous magnetization of a
Weiss ferromagnet.

The inverse Langevin function is one of the beneficiaries of Mez\"{o}, Baricz
\cite{[Mezo]} and Mugnaini's \cite{[Mugnaini]} results. More exactly: if
$L\left(  x\right)  =a,$ the function $L^{-1}\left(  a\right)  $ is:%

\begin{equation}
L^{-1}\left(  a\right)  =-2W\left(  \frac{2}{a+1};\frac{2}{a-1};\frac
{a-1}{a+1}\right) \label{63}%
\end{equation}

Also, it is a simple exercise to show that the inverse of the function
$L\left(  x\right)  /x$, denoted previously as $\lambda$, i.e. the solution in
$x$ of the equation $L\left(  x\right)  =\alpha x,$ is:%

\begin{equation}
x=\lambda\left(  \alpha\right)  =\label{64}%
\end{equation}

\[
=\frac{1}{2}W\left(  \frac{1+\sqrt{1-4\alpha}}{\alpha},\frac{1-\sqrt
{1-4\alpha}}{\alpha};-\frac{1+\sqrt{1-4\alpha}}{\alpha},-\frac{1-\sqrt
{1-4\alpha}}{\alpha};1\right)
\]

As, for the time being, there is no formula for the functions $W\left(
t_{1},t_{2};s_{1},s_{2};a\right)  ,$ eq. (\ref{64}) is of limited practical
use. However, the explicit form of $\lambda\left(  \alpha\right)  $\ was
obtained by Siewert and Burniston \cite{[SievBurnLangInv]}, but their result
is quite inconvenient for practical applications.

As the Brillouin functions can be written as a ratio of two polynomials, see
eq. (\ref{41}), they can be inverted by solving an algebraic equation. For
small values of $J,$ exact solutions can be obtained, as we shall see in
Subsection 5.2.

Concerning the function $\frac{B_{J}\left(  x\right)  }{x}$, the exact
expression of its inverse was obtained only for $J=1/2,$ i.e. \ for the
function $\frac{\tanh x}{x}.$ As $\tanh x$\ can be written in terms of one
exponential (for instance, of $e^{2x}$), the inverse of $\frac{\tanh x}{x}%
$\ is the inverse of the generalized Lambert function $W\left(  t;s;a\right)
$. This result is important in the context of ferromagnetism - and we shall
explain why. If we write the reduced magnetization in zero external field as
$m\left(  t\right)  ,$\ and define the function $\zeta\left(  t\right)  $\ by
(see \cite{[VB-VK]})%

\begin{equation}
m\left(  t\right)  =t\zeta\left(  t\right) \label{65}%
\end{equation}
the Weiss equation (eq. (10) of [9]) takes the form:%

\begin{equation}
\frac{\tanh\zeta\left(  t\right)  }{\zeta\left(  t\right)  }=t\label{66}%
\end{equation}
So, to invert the function $\left(  \tanh\zeta\right)  /\zeta$ means to obtain
the function $\zeta\left(  t\right)  $ and, consequently, the magnetization
$m\left(  t\right)  .$ Finally, $m\left(  t\right)  $\ can be written as:%

\begin{equation}
m\left(  t\right)  =t\zeta\left(  t\right)  =\frac{t}{2}W\left(  \frac{2}%
{t};-\frac{2}{t};-1\right) \label{67}%
\end{equation}
According to (\ref{59}):%

\begin{equation}
m\left(  t\right)  =1-2\sum_{n=1}^{\infty}\frac{L_{n}^{\prime}\left(
4n/t\right)  }{n}\left(  -e^{-2/t}\right)  ^{n}=\label{68}%
\end{equation}

\begin{equation}
=\frac{2}{t}+W_{\exp\left(  -2/t\right)  }\left(  -\frac{4}{t}\exp\left(
-2/t\right)  \right)  \ \label{69}%
\end{equation}
It is easy to see from eq. (\ref{68}) that the condition%

\begin{equation}
m\left(  0\right)  =1\label{70}%
\end{equation}
is fulfilled, as $L_{1}^{\prime}\left(  x\right)  =-1$ and $\lim
_{t\rightarrow0}e^{-2/t}=0.$ Due to the exponential term, the series in the
r.h.s of (\ref{68}) is rapidly convergent.

In the case of critical temperature, $t=1$ and the index of $W_{r}$ takes its
critical value (see eq. (\ref{62})), namely:%

\begin{equation}
r=\frac{1}{e^{2}}\label{71}%
\end{equation}
In this case:%

\begin{equation}
W\left(  2;-2;-1\right)  =2+W_{1/e^{2}}\left(  -\frac{4}{e^{2}}\right)
=0\label{72}%
\end{equation}
according to eq. (\ref{62}). Consequently, according to (\ref{69}), the
reduced magnetization at the critical temperature is zero:%

\begin{equation}
m\left(  1\right)  =0\label{73}%
\end{equation}

More than this, as mentioned just before eq. (\ref{62}), the magnetization is
not differentiable in $t=1,$ but is still continuos. This behavior is
compatible with the aspect of the experimental curve of reduced spontaneous
magnetization at critical temperature (see Fig. 2).

\begin{figure}[ptb]
\begin{center}
\includegraphics[width=\textwidth]{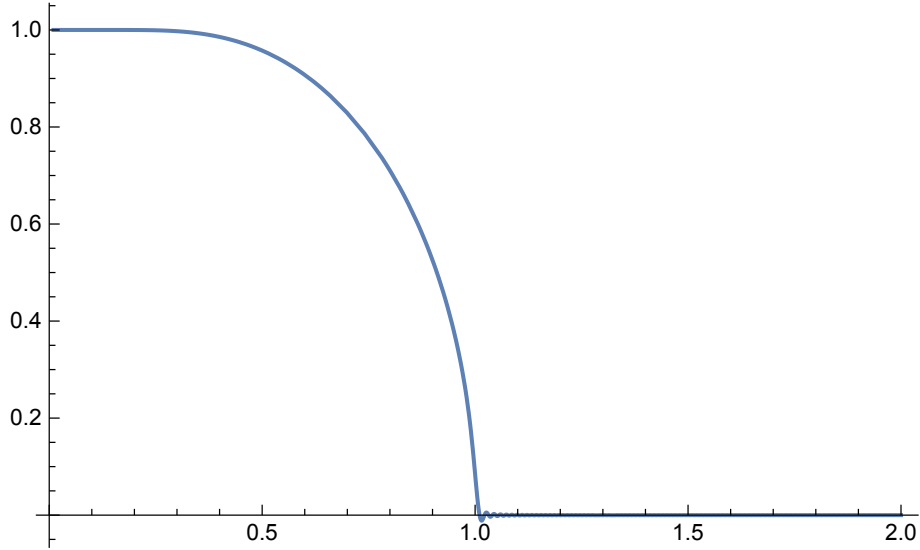}
\end{center}
\caption{The spontaneous magnetization $m(t)$, according to eq. (68), when the
first $n=1000$ terms of the series are took into account. If $n>2000$, the
oscillations for $t>1$ cannot be perceived by naked eye. }%
\end{figure}

If the magnetic field is non-zero, eq. (\ref{18}) for $J=1/2$ can be written as:%

\begin{equation}
\tanh\left(  \frac{m\left(  t,h\right)  }{t}+h\right)  =t\left(
\frac{m\left(  t,h\right)  }{t}+h\right)  -th\label{74}%
\end{equation}
or, putting%

\begin{equation}
u=\frac{m\left(  t,h\right)  }{t}+h\label{75}%
\end{equation}
as:%

\begin{equation}
\tanh u=t\left(  u-h\right) \label{76}%
\end{equation}
or, equivalently:%

\begin{equation}
e^{2u}\frac{u-h-\frac{1}{t}}{u-h+\frac{1}{t}}=-1\label{77}%
\end{equation}

According to (\ref{57}), (\ref{58}) the solution of this equation is:%

\begin{equation}
u=\frac{1}{2}W\left(  2h+\frac{2}{t};2h-\frac{2}{t};-1\right) \label{78}%
\end{equation}
so, finally we get:%

\begin{equation}
m\left(  t,h\right)  =\frac{t}{2}W\left(  2h+\frac{2}{t};2h-\frac{2}%
{t};-1\right)  -h\label{79}%
\end{equation}
or, using the $r-$Lambert function and eq. (\ref{61}):%

\begin{equation}
m\left(  t,h\right)  =W_{\exp\left(  -2h-2/t\right)  }\left(  \frac{4}%
{t}e^{-2h-2/t}\right)  -h\left(  1-t\right)  +1\label{80}%
\end{equation}

The critical isotherm is obtained making in the previous formula the
replacement $t\rightarrow1:$%

\begin{equation}
m\left(  1,h\right)  =W_{\exp\left(  -2\left(  h-1\right)  \right)  }\left(
4e^{-2\left(  h-1\right)  }\right)  +1\label{81}%
\end{equation}

For reasons explained in the previous section, these solution are quite
inconvenient for practical purposes. So, our next goal is to obtain an
alternative expression for $m\left(  t,h\right)  $, using simpler functions -
actually, a series expansion in powers of $h$, whose coefficients are known,
being expressed as elementary functions of $t.$

With:%

\begin{equation}
\zeta\left(  t,h\right)  =\frac{m\left(  t,h\right)  }{t},\ m\left(
t,h\right)  =t\zeta\left(  t,h\right) \label{82}%
\end{equation}
the equation of state (\ref{74}) becomes:%

\begin{equation}
t\zeta\left(  t,h\right)  =\tanh\left(  \zeta\left(  t,h\right)  +\frac{h}%
{t}\right) \label{83}%
\end{equation}

A closed form of $m\left(  t,0\right)  $ was obtained in the previous section,
eq. (\ref{69}), so an accurate analytic approximation of the function
$\zeta\left(  t,0\right)  $ is known, at least for low $t$. Differentiating
(\ref{83}) with respect to $h,$ we get:%

\[
t\frac{\partial\zeta\left(  t,h\right)  }{\partial h}=\left(  1-t^{2}\zeta
^{2}\left(  t,h\right)  \right)  \left(  \frac{\partial\zeta\left(
t,h\right)  }{\partial h}+\frac{1}{t}\right)
\]
and:%

\begin{equation}
\frac{\partial\zeta\left(  t,h\right)  }{\partial h}=\frac{1}{t}\frac{\left(
1-t^{2}\zeta^{2}\left(  t,h\right)  \right)  }{\left(  -1+t+t^{2}\zeta
^{2}\left(  t,h\right)  \right)  }\label{84}%
\end{equation}
Consequently:%

\begin{equation}
\left.  \frac{\partial\zeta\left(  t,h\right)  }{\partial h}\right\vert
_{h=0}=\frac{1}{t}\frac{\left(  1-t^{2}\zeta^{2}\left(  t,0\right)  \right)
}{\left(  -1+t+t^{2}\zeta^{2}\left(  t,0\right)  \right)  }\label{85}%
\end{equation}

So, the value of the previous expression is known, as the function
$\zeta\left(  t,0\right)  $ is known, e.g. via $m(t,0)$ provided by
(\ref{69}). This property of the derivative remains valid in any order;
consequently, we can write%

\begin{equation}
\zeta\left(  t,h\right)  =\zeta\left(  t,0\right)  +\left.  \frac
{\partial\zeta\left(  t,h\right)  }{\partial h}\right\vert _{h=0}h+\frac{1}%
{2}\left.  \frac{\partial^{2}\zeta\left(  t,h\right)  }{\partial h^{2}%
}\right\vert _{h=0}h^{2}+...\label{86}%
\end{equation}
In this way, we obtain a series expansion for $m\left(  t,h\right)  $,
according to (\ref{82}), valid for small values of $h.$ It should have in
mind, however, that we did not investigate the convergence of this series.

For $J>1/2,$ $\ $the inversion of $\frac{B_{J}\left(  x\right)  }{x}$ requests
the calculation of a root of the exponential polynomial (\ref{56}) containing
two different exponentials - a case not yet studied. However, for such
situations, accurate approximants have been proposed.

\subsection{Algebraic approach for obtaining $B_{J}^{-1}$}

As we already shown, we can put the Brillouin functions in an algebraic form,
eqs. (\ref{41}), and it is convenient to discuss separately the integer and
half-integer case.

If $J=n:$%

\begin{equation}
B_{n}\left(  z\right)  =\frac{1}{n}\frac{nz^{2n+2}-\left(  n+1\right)
z^{2n+1}+\left(  n+1\right)  z-n}{\left(  z^{1+2n}-1\right)  \left(
z-1\right)  }\label{87}%
\end{equation}
(the lower index of $z$ used in eqs. (\ref{43}) - (\ref{47}) has been dropped,
in order to avoid too complicated notations) and eq. (\ref{41}) gives the
following equation for $z\left(  t\right)  $, directly connected to the
inverse Brillouin function by (\ref{47}):%

\begin{equation}
n\left(  1-t\right)  z^{2n+2}-\left(  n\left(  1-t\right)  +1\right)
z^{2n+1}+\left(  n\left(  1+t\right)  +1\right)  z-n\left(  1+t\right)
=0\ ,\ t<1\label{88}%
\end{equation}
so, an equation of degree $2n+2=2\left(  J+1\right)  .$ The value of $z\left(
t\rightarrow1\right)  $ gives the asymptotic behavior of the root, important
for obtaining the asymptotic behavior of the inverse Brillouin function,
according to (\ref{51}). In order to find $z\left(  t\rightarrow1\right)  $,
we shall put $t=1$ in the coefficients of (\ref{88}), excepting the dominant
term (so, the substitution $t=1$\ does not affect the degree of the equation):%

\begin{equation}
n\left(  1-t\right)  z^{2n+2}-z^{2n+1}+\left(  2n+1\right)  z-2n=0\label{89}%
\end{equation}
Looking for a root of the form:%

\begin{equation}
z=\frac{a}{1-t}\label{90}%
\end{equation}
we get:%

\begin{equation}
z_{n}\left(  t\rightarrow1\right)  =\frac{1}{n\left(  1-t\right)  }\label{91}%
\end{equation}

It is easy to check that the result has the same form, for integer and
half-integer indices, so the asymptotic form of the inverse Brillouin function is:%

\begin{equation}
B_{J}^{-1}\left(  t\rightarrow1\right)  =J\ln z_{J}\left(  t\right)
=J\ln\frac{1}{J\left(  1-t\right)  }\rightarrow J\ln\frac{1}{1-t}\label{92}%
\end{equation}
as the constant term $-J\ln J$ is negligible, compared to the singular one.
The same result has been obtained by Kr\"{o}ger \cite{[Kroger]}, using a
different approach.

Denoting by $P_{2n+2}\left(  z;t\right)  $ the l.h.s. of (\ref{88}), we can
see that this polynomial has the following property:%

\begin{equation}
P_{2n+2}\left(  z;t\right)  =-z^{2n+2}P_{2n+2}\left(  \frac{1}{z};-t\right)
\label{93}%
\end{equation}

It is easy to check that%

\begin{equation}
\left[  \frac{\partial P_{2n+2}\left(  z;t\right)  }{\partial z}\right]
_{z=1}=0,\ \left[  \frac{\partial^{2}P_{2n+2}\left(  z;t\right)  }{\partial
z^{2}}\right]  _{z=1}=0,\ \left[  \frac{\partial^{3}P_{2n+2}\left(
z;t\right)  }{\partial z^{3}}\right]  _{z=1}\neq0\label{94}%
\end{equation}
so, $z=1$ is a double root of $P_{2n+2}\left(  z;t\right)  $. Removing this
root, the degree of equation (\ref{88}) becomes $2n$, but its tetranomic
character is lost.

Coming back to the polynomial $P_{2n+2}\left(  z,t\right)  $ , it contains
only four non-zero terms - so, (\ref{88}) is a tetranomial equation. Unlike
the trinomial case (containing three terms), where a general formula for the
roots is available \cite{[Passare]}, \cite{[Belkic]}, there is no such formula
for eq. (\ref{88}), at the best of author's knowledge, even if its form is
quite simple (the polynomial contains only the two highest degrees, i.e.
$z^{2n+2},\ z^{2n+1},$ and the two lowest degrees, $z^{1},z^{0}$ - the powers
of order $2,\ 3,\ ...\ 2n$ are missing); more than this, a quite interesting
symmetry relation is satisfied, as we just could see, eq. (\ref{93}). However,
as a general theory of the tetranomical equations is worked out
\cite{[Passare]}, it could produce exact (series expansion) results; this
happends, for example, for a tetranomical quintic equation.

It would be also possible, in principle, to use a Tschirnhaus transformation
(see \cite{[Adamchik]}, \cite{[VB-quint]}) in order to remove one of the
intermediate terms of the polynomial $P_{2n+2},$ and to reduce the tetranomic
(\ref{88}) to a trinomic.

So, we can expect that an analytical formula for the root $z\left(  t\right)
$ of (\ref{88}), could be also obtained.

Removing the double root of $P_{2n+1},$ eq. (\ref{88}) takes the form :%

\begin{equation}
\sum_{k=0}^{2n}\left(  k-\left(  t+1\right)  n\right)  z^{k}=0\label{95}%
\end{equation}

As the coefficients of the polynomial change sign only once, by virtue of
Descartes' criterion (\cite{[Wiki-Descartes]}), it has a unique positive real
root. It is localized in the interval $\left(  1,\infty\right)  $ and gives
the inverse Brillouin function, according to (\ref{47}). The algebraic form of
Brillouin functions was introduced and used for finding $B_{J}^{-1}$ by Millev
and Fahnle \cite{[MillevAmJPh]}, \cite{[Millev-pss-1992]},
\cite{[Millev-pss-1993]}. Even if the algebraic approach is useful, it cannot
provide a full solution for obtaining $B_{J}^{-1}$ for any $J$, and cannot
avoid the use of tranascendental equations for obtaining the explicit form of
the temperature dependence of the magnetization of a Weiss ferromagnet (to
mention the simplest form of the equation of state).

Similarly, from (\ref{42}):%

\begin{equation}
\overline{S}_{n}\left(  z\right)  =\frac{1}{z^{n}}\frac{z^{2n+1}-1}%
{z-1}=t\label{96}%
\end{equation}
we get the following equation for $\widetilde{z}\left(  t\right)  $:%

\begin{equation}
\widetilde{z}^{2n+1}-t\widetilde{z}^{n+1}+t\widetilde{z}^{n}-1=0\label{97}%
\end{equation}
so a tetranomic equation, of degree $2n+1=2J+1$ in $\widetilde{z}$. The
inverse function $S_{n}^{-1}\left(  t\right)  $ is obtained using eq.
(\ref{49}). The polynomial%

\begin{equation}
Q_{2n+1}\left(  \widetilde{z};t\right)  =\widetilde{z}^{2n+1}-t\widetilde
{z}^{n+1}+t\widetilde{z}^{n}-1\label{98}%
\end{equation}
obeys the symmetry relation:%

\begin{equation}
Q_{2n+1}\left(  \widetilde{z};t\right)  =-\widetilde{z}^{2n+1}Q_{2n+1}\left(
\frac{1}{\widetilde{z}};-t\right)  \label{99}%
\end{equation}
The polynomial $Q_{2n+1}\left(  \widetilde{z};t\right)  $ has one non-physical
root, $z=1,$ so, removing this root, the degree of eq. (\ref{99}) decreases to
$2n=J,$ but it is no more tetranomical.

If $J=n+\frac{1}{2}:$%

\begin{equation}
B_{n+\frac{1}{2}}\left(  z\right)  =\frac{1}{2n+1}\frac{\left(  2n+1\right)
z^{2n+3}\allowbreak-\left(  2n+3\right)  z^{2n+2}+\left(  2n+3\right)
z-\left(  2n+1\right)  }{\left(  z^{2n+2}-1\right)  \left(  z-1\right)
}\label{100}%
\end{equation}
and, from (\ref{41}), the inverse function $B_{n+\frac{1}{2}}^{-1}$\ is easily
obtained from the convenient root of the equation:%

\begin{equation}
\left(  2n+1\right)  \left(  1-t\right)  z^{2n+3}+\left(  -\left(
2n+3\right)  +t\left(  2n+1\right)  \right)  z^{2n+2}+\label{101}%
\end{equation}

\[
+\left(  \left(  2n+3\right)  +t\left(  2n+1\right)  \right)  z-\left(
2n+1\right)  \left(  1+t\right)  =0
\]
so, a tetranomical equation of degree $2n+3=2\left(  J+1\right)  $ in $z.$

The polynomial%

\[
\mathcal{P}_{2n+3}\left(  z;t\right)  =
\]

\begin{equation}
=\left(  2n+1\right)  \left(  1-t\right)  z^{2n+3}+\left(  -\left(
2n+3\right)  +t\left(  2n+1\right)  \right)  z^{2n+2}+\label{102}%
\end{equation}

\[
+\left(  \left(  2n+3\right)  +t\left(  2n+1\right)  \right)  z-\left(
2n+1\right)  \left(  1+t\right)
\]
has the symmetry property:%

\begin{equation}
\mathcal{P}_{2n+3}\left(  z;t\right)  =-z^{2n+3}\mathcal{P}_{2n+3}\left(
\frac{1}{z};-t\right)  \label{103}%
\end{equation}
and%

\[
\left[  \frac{\partial\mathcal{P}_{2n+2}\left(  z;t\right)  }{\partial
z}\right]  _{z=1}=0,\ \left[  \frac{\partial_{2n+2}^{2}\mathcal{P}\left(
z;t\right)  }{\partial z^{2}}\right]  _{z=1}=0,\
\]

\begin{equation}
\ \left[  \frac{\partial_{2n+2}^{3}\mathcal{P}\left(  z;t\right)  }{\partial
z^{3}}\right]  _{z=1}\neq0\label{104}%
\end{equation}

So, removing the unphysical double root, (\ref{101}) becomes an equation of
degree $2n+1=2J,\ $but it looses its tetranomic character.

Also:%

\begin{equation}
\overline{S}_{n+\frac{1}{2}}\left(  z\right)  =z^{n+\frac{1}{2}}+z^{n-\frac
{1}{2}}+...+z^{\frac{1}{2}}+z^{-\frac{1}{2}}+...+z^{-\left(  n-\frac{1}%
{2}\right)  }+z^{-\left(  n+\frac{1}{2}\right)  }\label{105}%
\end{equation}
With%

\begin{equation}
z=Z^{2},\ Z=e^{\frac{x}{2J}}\label{106}%
\end{equation}
we get:
\[
S_{n+\frac{1}{2}}\left(  Z\right)  =Z^{2n+1}+Z^{2n-1}+...+Z+\frac{1}%
{Z}+...+\frac{1}{Z^{2n-1}}+\frac{1}{Z^{2n+1}}=
\]

\begin{equation}
=\frac{Z^{2n+3}-1}{Z^{2n+1}\left(  Z-1\right)  }\ \label{107}%
\end{equation}
and the equation which gives $\widetilde{Z}\left(  t\right)  $, linked to the
inverse of $S_{n+\frac{1}{2}}$ according to (\ref{49}) is:%

\begin{equation}
\widetilde{Z}^{4n+4}-t\widetilde{Z}^{2n+3}+t\widetilde{Z}^{2n+1}%
-1=0\label{108}%
\end{equation}
so, a tetranomic equation of degree $4n+4=2\left(  2J+1\right)  $ in
$\widetilde{Z},$ or an equation of degree $2n+2=2J+1$ in $\widetilde
{U}=\widetilde{Z}+\frac{1}{\widetilde{Z}}$, which is no more tetranomic.

The polynomial%

\begin{equation}
\widetilde{Q}_{4n+4}\left(  x;t\right)  =x^{4n+4}-tx^{2n+3}+tx^{2n+1}%
-1\label{109}%
\end{equation}
has the symmetry property:%

\begin{equation}
\widetilde{Q}_{4n+4}\left(  x;t\right)  =-x^{4n+4}\widetilde{Q}_{4n+4}\left(
\frac{1}{x};-t\right)  \label{110}%
\end{equation}

Let us examine now, in some detail, the situation of these algebraic
equations, for small values of $J.$

For $J=1,$ (\ref{88}) gives:%

\begin{equation}
\left(  1-t\right)  z^{4}+\left(  t-2\right)  z^{3}+\left(  t+2\right)
z-\left(  1+t\right)  =0\label{111}%
\end{equation}
Noting that the l.h.s. of (\ref{111}) can be written as%

\begin{equation}
\left(  \left(  1-t\right)  z^{2}-tz-\left(  t+1\right)  \right)  \left(
z-1\right)  ^{2}\label{112}%
\end{equation}
and disregarding the unphysical root $t=1,$ we obtain:%

\begin{equation}
z\left(  t\right)  =\frac{t\pm\sqrt{t^{2}+4\left(  1-t^{2}\right)  }}{2\left(
1-t\right)  }\rightarrow\frac{t+\sqrt{4-3t^{2}}}{2\left(  1-t\right)
}\label{113}%
\end{equation}
a result also reported by Kr\"{o}ger, eq. (D.7) of \cite{[Kroger]}.

From (\ref{97}), we get:%

\begin{equation}
\overline{S}_{1}\left(  z\right)  =z+1+\frac{1}{z}\label{114}%
\end{equation}
so the inverse function $S_{1}^{-1}\left(  x\right)  $\ (see (\ref{49})) can
be easily obtained.

For $J=3/2,$ the inverse Brillouin function is a root $z\left(  t\right)
$\ of the equation:%

\begin{equation}
3\left(  1-t\right)  z^{3}+\left(  1-3t\right)  z^{2}-\left(  1+3t\right)
z-3\left(  1+t\right)  =0\label{115}%
\end{equation}
Its exact solution will be given in Subsection 5.3.

According to (\ref{105}), $S_{3/2}^{-1}\left(  t\right)  $ is given by the
root of the equation:%

\begin{equation}
S_{\frac{3}{2}}\left(  z\right)  =z^{\frac{3}{2}}+z^{\frac{1}{2}}+z^{-\frac
{1}{2}}+z^{-\frac{3}{2}}=t\label{116}%
\end{equation}
Putting%

\begin{equation}
z^{\frac{1}{2}}+z^{-\frac{1}{2}}=u\label{117}%
\end{equation}
(\ref{116}) takes the form%

\begin{equation}
u^{3}-2u-t=0\label{118}%
\end{equation}
so it is a cubic trinomial equation, with the physically convenient solution
given by (\cite{[Glasser]}, eq. (12b)):%

\begin{equation}
u\left(  t\right)  =-\frac{1}{\sqrt{3}}\sin\left(  \frac{1}{3}\arcsin\left(
-\frac{3\sqrt{3}}{4\sqrt{2}}t\right)  \right)  +\cos\left(  \frac{1}{3}%
\arcsin\left(  -\frac{3\sqrt{3}}{4\sqrt{2}}t\right)  \right)  \label{119}%
\end{equation}

For $J=2,$ the inverse Brillouin function is obtained from the root of the equation:%

\begin{equation}
\allowbreak2\left(  1-t\right)  z^{4}+\left(  1-2t\right)  z^{3}%
-2tz^{2}-\left(  1+2t\right)  \allowbreak z-2\left(  1+t\right)
=0,\ \ 0<t<1\label{120}%
\end{equation}
The exact solution of this equation will be given separately, in Subsection
5.4. However, $S_{2}^{-1}$ can be obtained solving the equation:%

\begin{equation}
S_{2}=z^{2}+\frac{1}{z^{2}}+z+\frac{1}{z}+1=t\label{121}%
\end{equation}
which is a second order equation in $z+\frac{1}{z}.$

For $J=5/2,$ the equation which generates $B_{5/2}^{-1}$ is quintic, with $t-
$dependent coefficients. It can be, in principle, solved exactly, but the
solution obtained in this way is of no practical use. However, the equation
which inverts $S_{\frac{5}{2}}\left(  z\right)  $ is much simpler:%

\begin{equation}
u^{5}-4u^{3}+3u=t,\ u=z+\frac{1}{z}\label{122}%
\end{equation}
as all the coefficients of the variables are numbers; it can be solved using
the method described in \cite{[VB-quint]}.

For $J=3,$ the tetranomic equation which gives $B_{3}^{-1}$ is sextic, so
practically unsolvable, but%

\begin{equation}
S_{3}=z^{3}+z^{2}+z+1+\frac{1}{z}+\frac{1}{z^{2}}+\frac{1}{z^{3}}\label{123}%
\end{equation}
is a quite simple cubic equation in $u=z+\frac{1}{z}\ $(the only $t-$
dependent term is the free one).

Consequently, for some cases, especially for $n=2$ and $n=3,$ it could be
useful to obtain $S_{n}^{-1}$ and then $B_{n}^{-1},$ using (\ref{37}).

\subsection{The exact expression of $B_{\frac{3}{2}}^{-1}$}

The inversion of the Brillouin function $B_{3/2}$ is done if we find the
convenient root of:%

\begin{equation}
\allowbreak3\left(  1-t\right)  z^{3}+\left(  1-3t\right)  z^{2}-\left(
1+3t\right)  z-3\left(  1+t\right)  =0\label{124}%
\end{equation}
$\allowbreak$

Following one of the standard approaches (see for instance Wolfram resources
\cite{[Wolfram]}\ or, more specifically, \cite{[VB-Cior]}), one finds that the
eq. (\ref{124}) has only one real root, namely:%

\begin{equation}
z\left(  t\right)  =\sqrt{\frac{4\left\vert p\right\vert }{3}}\cosh\left(
\frac{1}{3}\cosh^{-1}K\right)  -\frac{1}{3}a_{2},\ \ 0<t<1\label{125}%
\end{equation}
where we put (using standard notations for the roots of a cubic equation):%

\begin{equation}
\left\vert p\right\vert =\frac{2\left(  -9t^{2}+6t+5\right)  }{27\left(
1-t\right)  ^{2}},\ a_{2}=\frac{\left(  3t-1\right)  }{\allowbreak3\left(
t-1\right)  },\ K=\frac{1}{\sqrt{2}}\frac{135t^{3}-135t^{2}-171t+175}{\left(
5+6t-9t^{2}\right)  ^{3/2}}\label{126}%
\end{equation}
It is visible that the root has the form:%

\begin{equation}
z\left(  t\right)  =\frac{1}{1-t}\cdot\zeta\left(  t\right)  \label{127}%
\end{equation}
where $\zeta\left(  t\right)  $ is a regular function.

Finally,%

\begin{equation}
B_{3/2}^{-1}\left(  t\right)  =\frac{3}{2}\ln z\left(  t\right)  \label{128}%
\end{equation}
Its asymptotic behavior is:%

\begin{equation}
B_{3/2}^{-1}\left(  t\right)  =-\frac{3}{2}\ln\left(  1-t\right)  \label{129}%
\end{equation}
in accordance with (\ref{51}).

The roots (\ref{113}) and (\ref{125}) were firstly obtained by Millev and
F\"{a}hnle \cite{[Millev-pss-1993]}; actually, (\ref{125}) was written in
algebraic form.

\subsection{The exact expression of $B_{2}^{-1}$}

In order to obtain $B_{2}^{-1},$ we have to solve the quartic equation (see
(\ref{88})):%

\begin{equation}
\allowbreak2\left(  1-t\right)  z^{4}+\left(  1-2t\right)  z^{3}%
-2tz^{2}-\left(  1+2t\right)  \allowbreak z-2\left(  1+t\right)
=0,\label{130}%
\end{equation}

As we shall apply one of the standard approaches for solving the quartic,
given by Wolfram resources (\cite{[WolframQuart]}, eq. 34), we shall write the
eq. (\ref{130}) in the form:%

\begin{equation}
\allowbreak z^{4}+a_{3}z^{3}+a_{2}z^{2}+a_{1}\allowbreak z+a_{0}=0\label{131}%
\end{equation}
with:%

\begin{equation}
a_{3}=\frac{\left(  1-2t\right)  }{2\left(  1-t\right)  },\ a_{2}=-\frac
{t}{\left(  1-t\right)  },\ a_{1}=-\frac{\left(  2t+1\right)  }{2\left(
1-t\right)  },\ a_{0}=-\frac{\left(  t+1\right)  }{\left(  1-t\right)
}\label{132}%
\end{equation}

We associate to the quartic (\ref{131}) the resolvent (\cite{[WolframQuart]},
eq. 34):%

\begin{equation}
y^{3}-a_{2}y^{2}+\left(  a_{1}a_{3}-4a_{0}\right)  y+\left(  4a_{2}a_{0}%
-a_{1}^{2}-a_{3}^{2}a_{0}\right)  =0\label{133}%
\end{equation}
\qquad

Our first task is to obtain a real root of this resolvent. Actually, it has
only one real root, namely:%

\begin{equation}
y_{1}=\frac{1}{3\left(  1-t\right)  }\left(  \sqrt{5\left(  9-8t^{2}\right)
}\sinh\left(  \frac{1}{3}\sinh^{-1}\left(  -\frac{t\left(  27-20t^{2}\right)
}{\sqrt{5}\left(  9-8t^{2}\right)  ^{3/2}}\right)  \right)  -1\right)
\label{134}%
\end{equation}

The roots of the quartic (\ref{131}) are obtained as simple combinations of quantities like:%

\begin{equation}
R=\sqrt{\frac{1}{4}a_{3}^{2}-a_{2}+y_{1}}\label{135}%
\end{equation}

\begin{equation}
D=\sqrt{\frac{3}{4}a_{3}^{2}-R^{2}-2a_{2}+\frac{1}{4}\frac{\left(  4a_{3}%
a_{2}-8a_{1}-a_{3}^{3}\right)  }{R}}\label{136}%
\end{equation}

The convenient root of the quartic (\ref{131}) is find to be:%

\begin{equation}
z_{1}=-\frac{1}{4}a_{3}+\frac{1}{2}R+\frac{1}{2}D\sim\frac{1}{1-t}\label{137}%
\end{equation}

Although its expression is cumbersome, it is easy to check that it has the
correct asymptotic behavior:%

\begin{equation}
\left[  z_{1}\right]  _{asympt}\sim\frac{1}{1-t}\label{138}%
\end{equation}

\section{Approximate expressions of the inverse of $B_{J}\left(  x\right)  $
and $B_{J}\left(  x\right)  /x$}

\subsection{The case of \ functions $B_{J}\left(  x\right)  $}

There are several approximations for the inverse Brillouin functions, each of
them corresponds, in principle, to a certain purpose: it should be relevant
for teaching, or for a specific theory (for instance, hysteretic physics), or
for illustrating a Pad\'{e} approximation, etc. One of simplest and attractive
for pedagogical use is that of Arrott \cite{[Arrott]}. The author proposes a
very simple approximation for the Brillouin functions, noting that the expression:%

\begin{equation}
B\left(  x\right)  =\frac{x}{\sqrt{x^{2}+a^{2}}}=\frac{1}{a}x+O\left(
x^{3}\right)  \label{139}%
\end{equation}
behaves similar to $B_{J}\left(  x\right)  ,$ if $a=\left(  J+1\right)  /3J:$%

\begin{equation}
B_{J}\left(  x\right)  \simeq\frac{x}{\sqrt{x^{2}+a_{J}^{2}}},\ a_{J}%
=\frac{J+1}{3J}\label{140}%
\end{equation}

It can be easily inverted, so:%

\begin{equation}
B_{J}^{-1}\left(  x\right)  \simeq\frac{a_{J}x}{\sqrt{1-x^{2}}}\label{141}%
\end{equation}
The approximation is really excellent for $J=4:$%

\begin{equation}
B_{4}\left(  x\right)  \simeq\frac{x}{\sqrt{x^{2}+\left(  \frac{12}{5}\right)
^{2}}}\label{142}%
\end{equation}
(in eq. (9) of \cite{[Arrott]}, it is written, incorrectly, $5/12$ instead of
$12/5$), when the error $\varepsilon$ is between $-0.65$ and $0.3\%$. For
$J\neq4,$ the error is somewhat larger, $\sim1\%.$

In the context of hysteretic physics, Takacs \cite{[Takacs2016]} proposed, for
$B_{J},$ a very simple approximation:%

\begin{equation}
B_{J}\left(  x\right)  \simeq\tanh\left(  d_{J}x\right)  \label{143}%
\end{equation}
with%

\begin{equation}
d_{J}=\frac{1}{2.667J}+0.25\label{144}%
\end{equation}

The error of these approximants are much larger than ia Arrott's
case:$\ 3.4\%-6.3\%\ \left(  J=1\right)  ;\ 7\%-10\%\ \left(  J=3/2\right)
;\ 9.3-12.5\%\ \left(  J=2\right)  ;$\ $10-14\%\ \ \left(  J=5/2\right)  ;$
$13-15\%\ \left(  J=3\right)  ;\ 13.5-16.6\%\ \left(  J=7/2\right)  .$

Even if, through the formula (\ref{143}), the inverse Brillouin functions can
be easily approximated by elementary functions (arctanh), Takacs proposes
another variant:%

\begin{equation}
B_{J}^{-1}\left(  x\right)  \simeq\frac{axJ^{2}}{1-bx^{2}}\label{145}%
\end{equation}
with:%

\begin{equation}
a=\frac{0.5\left(  1+2J\right)  }{2J\left(  J-0.27\right)  }+\frac{0.1}{J^{2}%
};\ b=0.8\label{146}%
\end{equation}
but this formula is unacceptable, as $B_{J}^{-1}\left(  x\right)  <0$ for
$1-bx^{2}<0,$ i.e. for $\frac{1}{\sqrt{0.8}}<x<1.$

In another paper \cite{[Takacs2001]}, Takacs introduces a unique function%

\begin{equation}
B\left(  C_{1},C_{2};x\right)  =C_{1}\coth C_{1}x-C_{2}\coth C_{2}x\label{147}%
\end{equation}
which can generate both Brillouin and Langevin functions:%

\begin{equation}
B\left(  1,0;x\right)  \equiv\lim_{C_{2}\rightarrow0}B\left(  1,C_{2}%
;x\right)  =L\left(  x\right)  \label{148}%
\end{equation}
but the benefits of this remark remained unexploited.

Also, Takacs claims that the characteristic curves of ferromagnetism (mainly,
of hysteretic physics) can be deduced from and described by a simple
combination of linear and hyperbolic functions:%

\begin{equation}
T\left(  x\right)  =A_{0}x+B_{0}\tanh C_{0}x\label{149}%
\end{equation}
and the complexity of an analytic theory makes compulsory the use of very
severe approximations for $B_{J}$ and $B_{J}^{-1}.$ It is an elementary
exercise to show that the inverse of $T\left(  x\right)  $\ is a generalized
Lambert function:%

\begin{equation}
x\left(  T\right)  =\frac{1}{2C_{0}}W\left(  -\frac{2C_{0}\left(
B_{0}+T\right)  }{A_{0}},\frac{2C_{0}\left(  B_{0}+T\right)  }{A_{0}%
};-1\right)  \label{150}%
\end{equation}
but this formula seems to be of limited practical use.

In a recent paper \cite{[Kroger]}, Kr\"{o}ger developed a comprehensive
analysis of the inverse Langevin and Brillouin functions. One of his results
is the following:%

\begin{equation}
B_{J}^{-1}\left(  t\right)  =\frac{1}{2}\frac{15-11\left(  1-\varepsilon
\right)  \left(  1+2\varepsilon\right)  t^{2}}{5+10\varepsilon-\left(
1-\varepsilon\right)  \left[  5+11\varepsilon\left(  1+2\varepsilon\right)
\right]  t^{2}}\ln\frac{1+t}{1-t},\ \ \varepsilon=\frac{1}{2J}\label{151}%
\end{equation}

This is a very simple and precise formula (maximum relative errors from
$1.5\%$ for $J=3/2$ to $0.6\%$ for $J=5$ and to $0.35\%$ for $J=10$), with
increasing accuracy for larger indices. This accuracy can be, in principle,
increased again, replacing the quadratic polynomials in (\ref{151}) with
quartic ones, but the expression of approximants becomes quite complicated. As%

\begin{equation}
\frac{1}{2}\left(  \frac{15-11\left(  1-\varepsilon\right)  \left(
1+2\varepsilon\right)  t^{2}}{5+10\varepsilon-\left(  1-\varepsilon\right)
\left[  5+11\varepsilon\left(  1+2\varepsilon\right)  \right]  t^{2}}\right)
_{t=1}=J\label{152}%
\end{equation}
the asymptotic limit of (\ref{151}) is%

\begin{equation}
B_{J}^{-1}\left(  t\rightarrow1\right)  =J\ln\frac{1+t}{1-t}\label{153}%
\end{equation}
The only reason for maintaining the $1+t$ term in (\ref{153}) is to satisfy
the symmetry condition%

\begin{equation}
B_{J}^{-1}\left(  t\right)  =-B_{J}^{-1}\left(  -t\right)  .\label{154}%
\end{equation}

However, it seems to be more convenient to obtain the approximant of
$B_{J}^{-1}\left(  t\right)  $ in the first quadrant, i.e. for $t>0,$ without
any symmetry restrictions, and to define its value for $t<0$ using (\ref{29}).
So, a simpler asymptotic formula can be adopted \cite{[VB-subm]}:%

\begin{equation}
B_{J}^{-1}\left(  t\rightarrow1\right)  \rightarrow J\ln\frac{1}%
{1-t}\label{155}%
\end{equation}

As the symmetry constraint is released, it is convenient to replace the
rational function in (\ref{138}) by a polynomial, more specifically, to
replace (\ref{137}) by \cite{[VB-subm]}:%

\begin{equation}
B_{J}^{-1}\left(  t\right)  =\frac{3J}{1+J}P_{J}\left(  t\right)  \ln\left(
\frac{1}{1-t}\right)  ,\label{156}%
\end{equation}
with the polynomial $P_{J}\left(  t\right)  $\ satisfying the conditions%
\begin{equation}
P_{J}\left(  0\right)  =1,\ P_{J}\left(  1\right)  =\frac{1+J}{3}\label{157}%
\end{equation}
as one can easily get from the properties of $B_{J}^{-1}.$

Similarly, one can obtain as many $\left(  t_{j},P_{J}\left(  t_{j}\right)
\right)  $ numerical pairs as we want, to be used in order to obtain the
explicit form of the polynomial $P_{J}\left(  t\right)  $ using the
\textit{Fitting curves to data} command in Mathematica. Choosing a
sufficiently large number of points $\left(  y_{j},P_{J}\left(  y_{j}\right)
\right)  ,$ we can obtain a sufficiently high accuracy, according to our
specific goal.

The expression of the approximant (\ref{156}) can be obtained directly using
the following code, proposed by Kr\"{o}ger (see ref. (10) in [28]):
\begin{verbatim}
Brillouin[J_][x_]:=(S+1)/S Coth[(S+1)/S x]-1/S Coth[x/S]/.S->2 J;

InverseBrillouinApproximant[J_][y_]:=Module[{x,Y,B,P,DATA},
  P[s_][Y_]:=(1+s)/(3 s)(x/.FindRoot[Brillouin[s][x]-Y,{x,0.9 Y}])/Log[1/(1-Y)];
  P[s_][1]:=(1+s)/3;
  P[s_][0]:=1;
  DATA=Table[{Y,P[J][Y]},{Y,{0,0.1,0.2,0.3,0.4,0.5,0.6,0.7,0.8,0.9,0.95,1}}];
  Fit[DATA,Table[Y^n,{n,0,30}],Y]3 J/(1+J)Log[1/(1-Y)]/.Y:>y];

\end{verbatim}

Evidently, the maximum value of $n$ in this code (here: 30) can be adapted to
the specific problem under examination.

The main advantages of this method are the accuracy, and the availability of
an analytical formula, for each $J$ separately. Actually, its accuracy can be
increased as much as needed, increasing the number of intermediate points
$y_{j}.$ Its main disadvantages might be that (1) the specific expression of
the polynomial $P_{J}\left(  y\right)  $ changes if the distribution of points
$\left(  y_{j},P_{J}\left(  y_{j}\right)  \right)  $ changes and (2) the
degree of the polynomial which grants a good accuracy is too large. Both
problems were studied by Tolea \cite{[Mugurel]}, who demonstrated that (1) if
the number of intermediate points is about $10^{6},$ the first 4 digits of the
polynomial coefficients do not change if this number increases and (2) a
polynomial as simple as a sextic one is sufficient to obtain an approximant
with an accuracy better than any of the approximants currently in use.

\subsection{The case of \ functions $B_{J}\left(  x\right)  /x$}

We shall consider firstly the case $J=1/2,$ the only one when there is an
analitic formula for the inverse of $B_{J}\left(  x\right)  /x.$ The inverse
of $B_{1/2}\left(  x\right)  /x=\tanh x/x,$ denoted $\zeta\left(  t\right)  $
in eq. (\ref{65}), is simply connected to the magnetization, $m\left(
t\right)  =t\zeta\left(  t\right)  ,$ a quantity with a direct physical
meaning. So, in order to avoid unnecessary complications, we shall discuss
here the approximants of the magnetization in a Weiss model, firstly for
$J=1/2.$ As $m\left(  t\right)  $ is the solution of the transcendental equation:%

\begin{equation}
m\left(  t\right)  =\tanh\frac{m\left(  t\right)  }{t}\label{158}%
\end{equation}
which is not analytic near $t=0$ and $t=1,$ we shall firstly find its
behaviour in the neighborhood of these points. Such an approach is
systematically used, for instance, in order to find a physically acceptable
solution of the Schroedinger equation. A wellknown example might be the
following: when we are solving the Schroedinger equation in a Coulombian
field, we are looking for a radial solution which "behaves well" near origin
and at infinity.

As shown in \cite{[VB-VK]}, eq. (30):%

\begin{equation}
m\left(  t\gtrsim0\right)  \simeq1+\frac{t}{2}W\left(  -\frac{4}{t}\exp\left(
-\frac{2}{t}\right)  \right)  \simeq1-2\exp\left(  -\frac{2}{t}\right)
\label{159}%
\end{equation}

Near $t=1,$ the non-analitic behaviour is familiar from Landau theory of phase transitions:%

\begin{equation}
m\left(  t\gtrsim0\right)  \sim\sqrt{1-t}\label{160}%
\end{equation}

So, on the whole interval $\left(  0,1\right)  $ it seems convenient to write
the magnetization in the form (\cite{[VB-VK]}, eq. (34)):%

\begin{equation}
m\left(  t\right)  =\left(  1-2\exp\left(  -\frac{2}{t}\right)  \right)
P_{1/2}\left(  t\right)  \sqrt{1-t}\ \label{161}%
\end{equation}
where the index $1/2$ in the polynomial $P_{1/2}\left(  t\right)  $ reminds us
that we are in the case $J=1/2;$ the index was dropped however in the notation
$m\left(  t\right)  $, for sake of simplicity. Clearly, this equation implies that:%

\begin{equation}
P_{1/2}\left(  0\right)  =1\label{162}%
\end{equation}

So, the polynomial $P_{1/2}\left(  t\right)  $ has the expression:%

\begin{equation}
P_{1/2}\left(  t\right)  =1+a_{1}t+...+a_{n}t^{n}=\frac{m\left(  t\right)
}{\left(  1-2\exp\left(  -\frac{2}{t}\right)  \right)  \sqrt{1-t}}\label{163}%
\end{equation}

The coefficients $a_{1},...a_{n}$ can be obtained by solving a linear system
of $n$ equations, obtained from the previous relation for $n$ couples of
values $\left(  t_{0},m\left(  t_{0}\right)  \right)  ;$ for a given $t_{0},$
the corresponding $m\left(  t_{0}\right)  $ is calculated solving numerically
the equation:%

\begin{equation}
m\left(  t_{0}\right)  =\tanh\frac{m\left(  t_{0}\right)  }{t_{0}}\label{164}%
\end{equation}

For $n=7,$ such a polynomial is obtained in [9], eq. (38). The deviation of
the approximate magnetization, obtaining introducing this polynomial in eq.
(\ref{147}), and the exact one, is less than $3\cdot10^{-3},$ for $0\lesssim
t\lesssim1/2$ and less than $3\cdot10^{-4}$ for $1/2\lesssim t\lesssim1.$

For applications in experimental physics - e.g. for obtaining the critical
temperature $T_{c}$ from experimental measurement of magnetization - a very
precise analytical of the magnetization for $t$ close to zero is useless, at
least because it does not behaves according to (\ref{161}), but according to
Bloch's law (see for instance \cite{[Kittel]}), so it is reasonable to replace
(\ref{161}) by:%

\begin{equation}
m_{J}\left(  t\right)  =Q_{J}\left(  t\right)  \sqrt{1-t}\label{165}%
\end{equation}
where the value of spin, $J$, was explicitely introduced. A number of five
polynomials of degree 7, $Q_{J}\left(  t\right)  ,$ for
$J=1/2,\ 1,\ ...5/2\ ,$ with relative deviations ov about $10^{-7},$ are given
in \cite{[VB-VK]}.

\section{Conclusions}

In this paper, we reviewed the results concerning the inverse Langevin and
Brillouin functions $L\left(  x\right)  $ and $B_{J}\left(  x\right)  $,
respectively, obtained recently by researchers working in several domains of
physics - rubber elasticity, rheology, solar energy conversion,
ferromagnetism, superparamagnetism, hysteretic physics - and of mathematics -
theory of transcendental or algebraic equations -, and also added some of
ours, new and yet unpublished. This review might be of interest, as
researchers working in so diverse fields are not necessarily aware of the
progress registered by their colleagues, focused on different physical
problems, which share, however, a common mathematical basis.

We put in value the physical significance of Langevin and Brillouin functions,
their similarities and differences. We explained how the inverses of $L\left(
x\right)  $ and $B_{J}\left(  x\right)  /x$ can be obtained using the recent
progress in the theory of generalized Lambert functions, and also presented
several approximants of these functions, interesting for applied physics. We
discussed the accuracy and usefulness of these approximants, from several
perspectives (pedagogical, for hysteretic physics, for ferromagnetism). We
also gave the exact expressions of $B_{3/2}^{-1}\left(  x\right)  $ and
$B_{2}^{-1}\left(  x\right)  ;$ even if they might be too complicated for
practical applications, they facilitate the understanding of general
properties of inverse Brillouin functions of arbitrary index, for instance
their asymptotic behavior.

This review, mainly focused on theoretical papers, might interest also the
experimentalists, as the mathematics was maintained at an accessible level.%

\[
\]

\textit{Conflict of interest disclosure: } The author declares that there is
no conflict of interest regarding the publication of this paper.

\begin{acknowledgement}
\bigskip The author is grateful to Dr. Mugurel Tolea, for illuminating
discussions. The financial support of the ANCSI - IFIN-HH project PN 18 09 01
01/2018 is also acknowledged.
\end{acknowledgement}

\end{document}